\documentclass[a4paper]{jpconf}
\usepackage{graphicx}
\newcommand{\mssm}{{\sc mssm}}
\newcommand{\susy}{{\sc susy}}
\newcommand{\msugra}{{\sc msugra}}
\newcommand{\qcd}{{\sc qcd}}
\newcommand{\sqcd}{{\sc susy-qcd}}
\newcommand{\lhc}{{\sc lhc}}
\newcommand{\dreg}{{\sc dreg}}
\newcommand{\dred}{{\sc dred}}
\newcommand{\mh}{\ensuremath{M_h}}
\newcommand{\mt}{\ensuremath{M_t}}
\newcommand{\mst}[1]{\ensuremath{M_{\tilde{t_{#1}}}}}
\newcommand{\msq}{\ensuremath{M_{\tilde q}}}
\newcommand{\mgl}{\ensuremath{M_{\tilde g}}}
\newcommand{\drbar}{\ensuremath{\overline{\textmd{\textsc{dr}}}}}

\newcommand{\alphas}{\ensuremath{\alpha_s}}
\newcommand{\alphat}{\ensuremath{\alpha_t}}

\newcommand{\epscalars}{\ensuremath{\varepsilon}-scalars}

\newcommand{\order}[1]{\ensuremath{{\cal O}\left( {#1} \right)}}
\newcommand{\orderat}{\order{\alphat}}
\newcommand{\orderatas}{\order{\alphat\alphas}}
\newcommand{\orderatasas}{\order{\alphat\alphas^2}}

\newcommand{\QGRAF}{\textmd{\textsc{qgraf}}}
\newcommand{\QTOE}{\textmd{\textsc{q2e}/\textsc{exp}}}
\newcommand{\FORM}{\textmd{\textsc{form}}}
\newcommand{\MATAD}{\textmd{\textsc{matad}}}

\newcommand{\hthreem}{{\sc H3m}}
\newcommand{\Mathematica}{{\sc Mathematica}}
\newcommand{\FeynHiggs}{{\sc FeynHiggs}}
\newcommand{\suspect}{{\sc SuSpect}}
\newcommand{\spheno}{{\sc SPheno}}
\newcommand{\softsusy}{{\sc softsusy}}
\newcommand{\slha}{{\sc slha}}
\newcommand{\tsil}{{\sc tsil}}

\begin{document}
\pagestyle{plain}
\title{%
  \vskip-4\baselineskip%
  {\normalsize HU-EP-11/55\\%
    \normalsize SFB/CPP-11-71}%
  \vskip-3\baselineskip%
  \vskip4\baselineskip
Three-Loop Calculation of the Higgs Boson Mass in
  Supersymmetry}
\author{Philipp Kant}
\address{Humboldt-Universit\"at zu Berlin}
\ead{philipp.kant@physik.hu-berlin.de}

\begin{abstract}
A Key feature of the minimal supersymmetric extension of the Standard
Model (\mssm{}) is the existence of a light Higgs boson, the mass of
which is not a free parameter but an observable that can be predicted
from the theory.  Given that the \lhc{} is able to measure the mass of a
light Higgs with very good accuracy, a lot of effort has been put into
a precise theoretical prediction.

We present a calculation of the \sqcd{} corrections to this
observable to three-loop order.  We perform multiple asymptotic
expansions in order to deal with the multi-scale three-loop diagrams,
making heavy use of computer algebra and keeping a keen eye on the
numerical error introduced.

We provide a computer code in the form of a Mathematica package that
combines our three-loop \sqcd{} calculation with the literature of
one- and two-loop corrections to the Higgs mass, providing a
state-of-the-art prediction for this important observable.
\end{abstract}

\section{Introduction}
The minimal supersymmetric extension of the Standard Model (\mssm{})
is a promising candidate for physics beyond the Standard Model.  Its
Higgs sector is a two-Higgs doublet model with the additional
constraint that supersymmetry relates the quartic Higgs couplings to
the gauge couplings of the theory.  This increases the predictiveness
of the model and allows the Higgs sector to be parametrised by just
two new parameters, the mass $M_A$ of the pseudoscalar Higgs and the
ratio $\tan\beta$ of the vacuum expectation values of the Higgs
doublets.

In particular, the mass \mh{} of the light Higgs boson is not a free
parameter, but can be predicted.  At tree level, \mh{} is bounded
above by $M_Z$, but radiative corrections shift the value
significantly.  Since \mh{} will be a precision observable once the
Higgs is found at the Large Hadron Collider (\lhc), it is imperative
to have a precise theoretical prediction.  Consequently, a lot of
effort has been put into the calculation of radiative corrections to
\mh{} at the one- and two-loop level (see, for example~\cite{%
  Hempfling:1993qq
  ,Zhang:1998bm,Heinemeyer:1998jw,%
  Pilaftsis:1999qt,Carena:2000dp,Espinosa:2001mm,%
  Degrassi:2001yf,Martin:2002wn}).  The remaining theoretical
uncertainty has been estimated to be about
$3-5\,$GeV~\cite{Degrassi:2002fi,Allanach:2004rh}, which is confirmed
by a study of the leading and next-to-leading terms in
$\ln(M_{SUSY}/M_t)$, where $M_{SUSY}$ is the typical scale of \susy{}
particle masses, at three-loop order~\cite{Martin:2007pg}.

This uncertainty of the theoretical prediction justifies a calculation
of the next term in the perturbative expansion.  A study of the
corrections to \mh{} shows that the contributions from loops of top
quarks and their superpartners, the stops, are dominant at the one-
and two-loop level.  In~\cite{Harlander:2008ju,Kant:2010tf}, we have
calculated three-loop \sqcd{} corrections to these diagrams.  The
calculation of these terms is the subject of this talk.

\section{Calculation of the Three-Loop Corrections}
Motivated by the discussion above, we restrict the calculation at the 
three-loop level to the \sqcd{} corrections where the Higgs couples to
a top quark or its superpartners.  In the perturbative expansion of
\mh{}, these are the terms of \orderatasas, where
\alphas{} is the strong coupling and \alphat{} denotes the coupling of
the Higgs to the top quark.

We thus have to evaluate three-loop propagator diagrams in \sqcd,
which faces us with two complications.  First, we need a convenient
regulator that respects supersymmetry, and second, a lot of masses
appear in our diagrams.

\subsection{Regularisation by Dimensional Reduction}
The regularisation workhorse of multi-loop calculations is dimensional
regularisation
(\dreg{})~\cite{Wilson:1972cf
},
where the number of spacetime dimensions is altered from $4$ to
$d=4-2\varepsilon$, and the divergences of the loop integrals are
manifest as poles in $\varepsilon$.

Unfortunately, \dreg{} does not respect supersymmetry.  An easy way to
see this is that changing the number of spacetime dimensions also
changes the number of degrees of freedoms of the vector fields, and
supersymmetry requires an equal number of bosonic and fermionic
degrees of freedom.  While it is possible (but tedious) to manually
construct finite counterterms that restore the supersymmetric ward
identities, a more convenient approach was suggested by Siegel under
the name of dimensional reduction (\dred)~\cite{Siegel:1979wq}.  The
main idea is that the change of dimensions need only affect momenta in
order to regularise the loop integrals.  In \dred{}, one splits the
four dimensional space into orthogonal spaces of dimension
$2\varepsilon$ and $d=4-2\varepsilon$, and restricts the loop momenta
to the $d$ dimensional subspace while all gauge fields are kept four
dimensional.  The $2\varepsilon$ components of the gauge fields
transform as a $2\varepsilon$ tuple of scalar fields and are called
\epscalars{}.

The consistent formulation of \dred{} is more involved than sketched
here~\cite{Stockinger:2005gx}, and the question whether the inclusion
of the \epscalars{} truly restores the supersymmetric ward identities
in all cases is not yet resolved.  However, \dred{} has successfully
been applied in many multi-loop
calculations\cite{Harlander:2006xq,Marquard:2007uj,Bednyakov:2007vm,Jack:2007ni,Signer:2008va,Mihaila:2009bn,Pak:2010cu,Harlander:2010wr,Kilgore:2011ta,Hermann:2011ha},
so far without observing a violation of the ward identities.

\subsection{Asymptotic Expansion in the Masses}
\label{sec:exp}
The difficulty of calculating a Feynman diagram rises with the number
of loops, the number of external momenta and the number of masses in
the diagram.  To obtain the \orderatasas{} terms of \mh, we have to
evaluate diagrams with three loops, two legs and a lot of masses:
treating the light quarks as massless and their superpartners as mass
degenerate with mass \msq, we are left with five masses: the top quark
mass \mt, the masses of its superpartners \mst{1}, \mst{2}, the mass
of the gluino \mgl{} and \msq.  Calculating these diagrams without any
approximations is not feasible with contemporary methods.

Given the mild dependence of the one-loop corrections to \mh{} on the
external momentum, it is reasonable to expand the diagrams in small
external momentum, reducing the problem to the evaluation of vacuum
diagrams.  A further simplification is possible by performing nested
asymptotic
expansions~\cite{Tkachov:1991ev
,Gorishnii:1986gn,Smirnov:1993ks
,Smirnov:1990rz
},
which expresses the multi-mass diagrams as a series in ratios of the
masses and logarithms of these ratios.  The coefficients in the series
involve only one-scale integrals.  Assuming that the ratios of the
masses are small, the series should give good approximations to the
original diagrams.

Of course, before carrying out the asymptotic expansions, one has to
identify which mass ratios are small.  Given that so far none of the
superpartner masses are known, this is an undecidable problem.  So,
instead of calculating with a fixed, known, hierarchy among the
masses, one has to systematically consider various possible mass
hierarchies and carry out the calculation for each of these.  To get a
prediction of \mh{} for a specific point in the parameter space of the
\mssm{}, where the masses of the superpartners have specific values,
one has to choose whichever hierarchy fits these values best.

We carry out the calculation using the following setup: In a first
step, all diagrams contributing to \mh{} at \orderatasas{} are found
using the program \QGRAF~\cite{Nogueira:1991ex}.  There are $30.717$
diagrams.  For each hierarchy, these diagrams are expanded
asymptotically using
\QTOE{}~\cite{Seidensticker:1999bb
}.  The one-scale
integrals are then calculated using the
\FORM{}~\cite{Vermaseren:2000nd} program
\MATAD{}~\cite{Steinhauser:2000ry}.

\subsection{Renormalisation}
\begin{figure}
  \includegraphics[width=\textwidth]{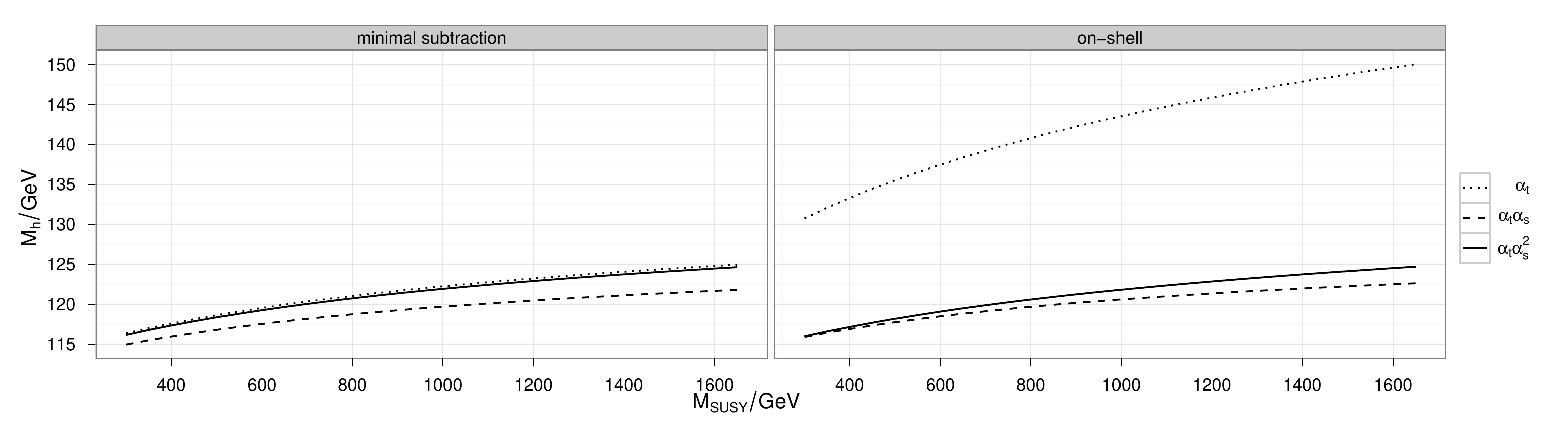}
  \caption{Behaviour of the perturbative series for \mh{} using the
    on-shell and the minimal subtraction scheme (for degenerate
    superpartner masses).  The short-dashed, long-dashed and solid
    lines are the one-, two-, and three-loop approximation to \mh{},
    respectively.  In the on-shell scheme, the distance between one-
    and two-loop approximation is much larger than in the \drbar{}
    scheme.}
\label{fig:schemedep}
\end{figure}

For the renormalisation of the parameters entering our calculation, we
adopt a variation of the \drbar{}-scheme, i.e. minimal subtraction
using dimensional reduction.  This leads to a much better behaviour of
the perturbative series compared to using on-shell renormalisation
(see Fig.~\ref{fig:schemedep}).  

\section{Computer Code}
In order to provide a precise prediction for the value of \mh{} within
the \mssm{}, our \orderatasas{} terms have to be combined with the
wealth of corrections from other sectors of the \mssm{} at one- and
two-loop level that are available in the literature~\cite{%
  Hempfling:1993qq
  ,Zhang:1998bm,Heinemeyer:1998jw,%
  Pilaftsis:1999qt,Carena:2000dp,Espinosa:2001mm,%
  Degrassi:2001yf,Martin:2002wn}.
Also, the choice of hierarchy mentioned in \ref{sec:exp} should be
automatised.  In~\cite{Kant:2010tf}, we have presented the
\Mathematica{} package \hthreem{}, which is publicly available, to
address these points.  For convenience, it implements the Susy Les
Houches Accord \slha{}~\cite{Skands:2003cj} and has an interface to
the spectrum generators \softsusy~\cite{Allanach:2001kg},
\suspect~\cite{Djouadi:2002ze} and
\spheno~\cite{Porod:2011nf
}.

The program uses
\FeynHiggs{}~\cite{Heinemeyer:1998yj
,Degrassi:2002fi
}
to get a two-loop approximation of \mh.  An obstacle for adding our
terms of \orderatasas{} to the result delivered by \FeynHiggs{} is the
usage of on-shell renormalisation in FeynHiggs.  In order to be
consistent, we have to convert the \orderat{} and \orderatas{} terms
that contain on-shell parameters to the \drbar{} scheme.  Thanks
to~\cite{Degrassi:2001yf}, which gives a compact expression for
these terms both in the \drbar{} and in the on-shell scheme, this
obstacle is easily overcome.

The corrections to \mh{} depend very strongly on the mass \mt{} of the
top quark and the strong coupling \alphas.  These parameters must be
known, within \sqcd{} renormalised in the \drbar{} scheme, as
precisely as possible.  In~\cite{Martin:2005ch}, the relation between
the top mass in the \drbar- and the on-shell scheme in \sqcd{} has
been calculated to \order{\alphas^2}.  Using the library
\tsil~\cite{Martin:2005qm}, this relation can be used to obtain
$\mt^{\drbar}$ from the known value of the top quark pole mass.

We determine \alphas{} in \sqcd{} from the value of \alphas{} in
five-flavour \qcd{} at the $Z$ mass
following~\cite{Harlander:2007wh} by running, within five-flavour
\qcd, to the decoupling scale where we perform the transition to the
full theory.  We then run, within \sqcd{}, to the desired value of the
renormalisation scale.

In order to choose the most appropriate hierarchy and get an estimate
for the error introduced by the asymptotic expansion, we compute an
approximation to the two-loop corrections of \orderatas{} using
asymptotic expansions in the mass ratios and compare this
approximation to the result of~\cite{Degrassi:2001yf}.  The hierarchy
that matches the exact two-loop result best is chosen for the
calculation of the three-loop term, and the error at two-loop level is
recorded to get a handle on the error in the three-loop result (see
Fig.~\ref{fig:experror}).

\section{Numerical Results}
\begin{figure}
  \includegraphics[angle=90,width=\textwidth]{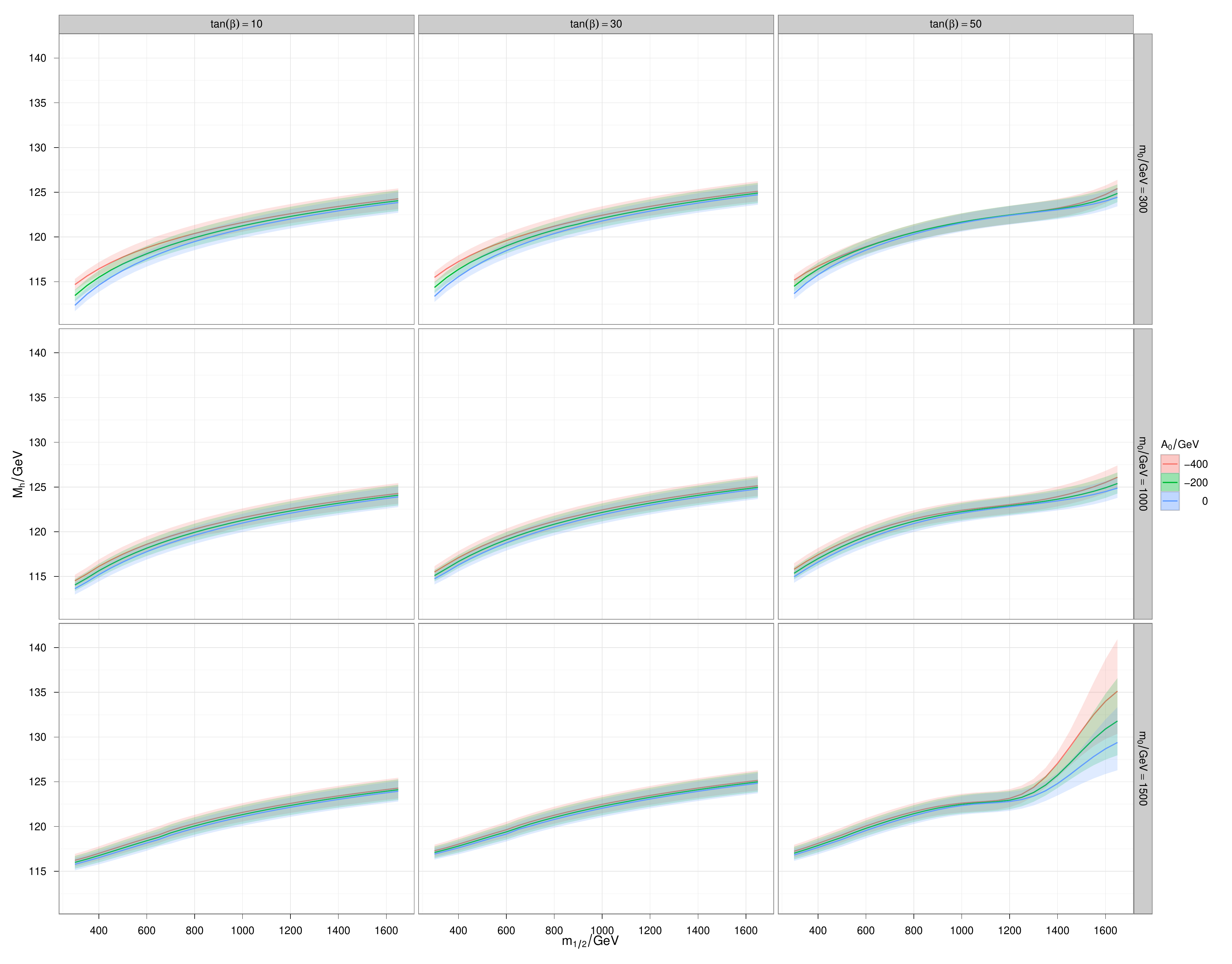}
  \caption{Theoretical prediction for the light Higgs mass in the
    \msugra{} model.  We plot \mh{} over the fermion mass parameter
    $m_{1/2}$.  The panels show, from left to right, an increasing
    value of $\tan\beta$, and, from top to bottom, of the scalar mass
    parameter $m_0$.  We show three curves according to different
    values of the trilinear coupling $A_0$.}
  \label{fig:mhmsugra}
\end{figure}

In this chapter, we present a state-of-the-art numerical prediction
for \mh{} in the \msugra{} model that has been obtained with
\hthreem.  Fig.~\ref{fig:mhmsugra} shows the dependence of \mh{} on
the parameters $m_{1/2}$, $m_0$, $\tan\beta$, and $A_0$ for $\mu>0$.
The shaded bands around the individual curve show the variation of
\mh{} when \mt{} is varied between $171.4\,$GeV and $174.4\,$GeV.  We
restrict the plot to values of $m_{1/2}>300\,$GeV in light of the
latest exclusions from the \lhc{} experiments~\cite{Aad:2011qa,:2011iu,Padhi:2011rs,Schettler:2011hc}.

We observe that of all \msugra{} parameters, varying $m_{1/2}$ has by
far the largest impact on the Higgs mass.  We also observe that the
present uncertainty of the top mass directly translates to a
parametric uncertainty of about one GeV of the Higgs mass.

\begin{figure}
  \includegraphics[width=\textwidth]{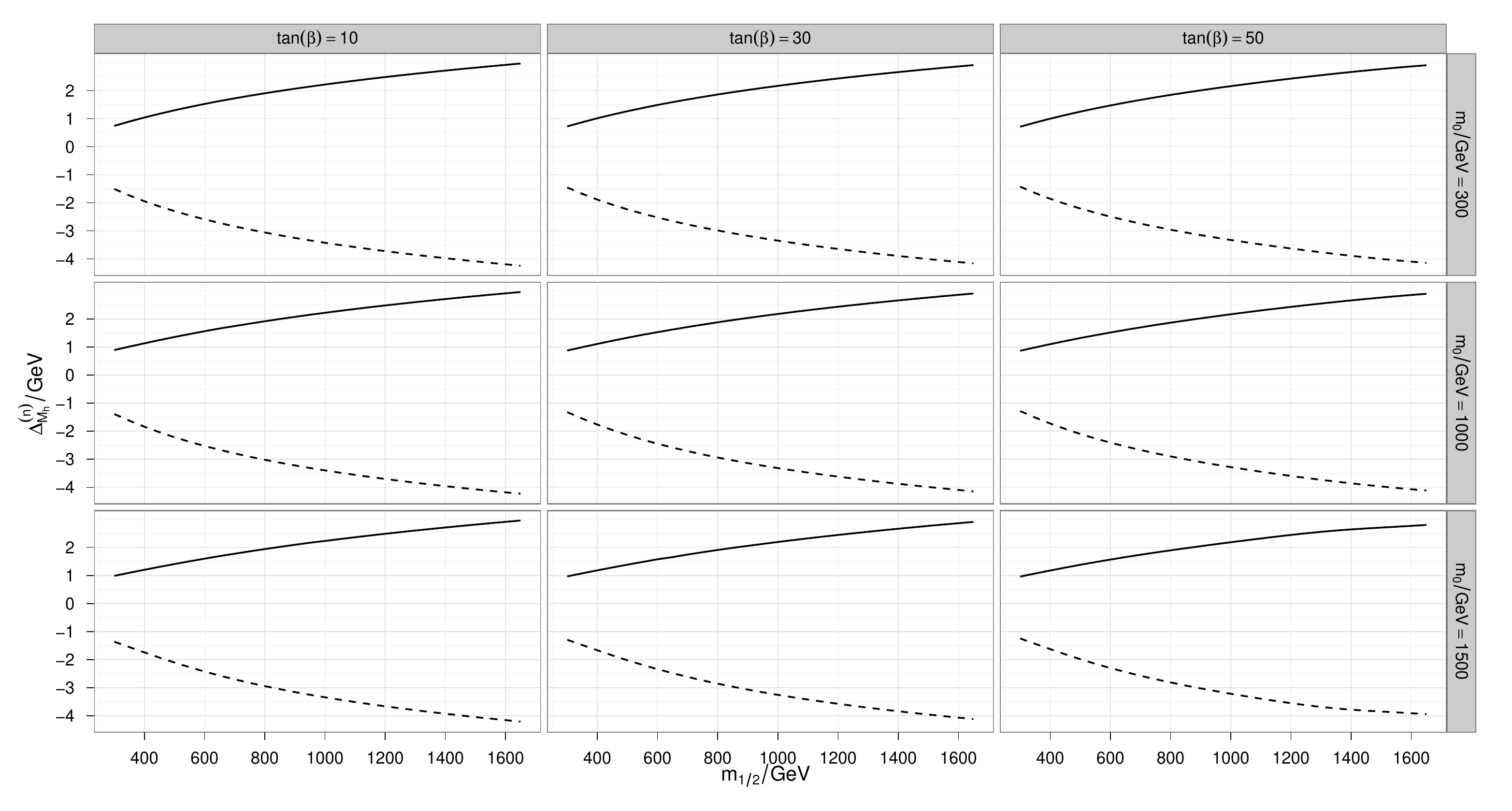}
  \caption{Two-loop (dashed) and three-loop (solid) contributions to
    \mh{}.  The same conventions as in Fig.~\ref{fig:mhmsugra} are
    adopted.  For lucidity, only the curves for $A_0=0$ and
    $\mt=172.9\,$GeV are shown.  
    We observe that the three-loop terms are smaller than the two-loop
    terms, but not by a large factor.  This suggests a conservative
    approach when estimating higher order effects.}
  \label{fig:dmh2l3l}
\end{figure}

To estimate the effect of unknown higher orders,
Fig.~\ref{fig:dmh2l3l} shows the two- and three-loop corrections to
\mh.  We observe a partial cancellation between the two- and
three-loop correction.  We also observe that while the higher order
corrections do decrease in magnitude, they do not do so by a large
factor.  Thus, we should be careful when estimating the magnitude of
the missing higher order contributions.  We choose to assign $50\,$\%
of the three-loop contribution as a theoretical error.  This amounts
to a hundred MeV for low values of $m_{1/2}$ and to about one GeV
for $m_{1/2}$ above one GeV.  Using the scale variation as an error
estimate would lead to a smaller error~\cite{Kant:2010tf}.

\begin{figure}
  \includegraphics[width=\textwidth]{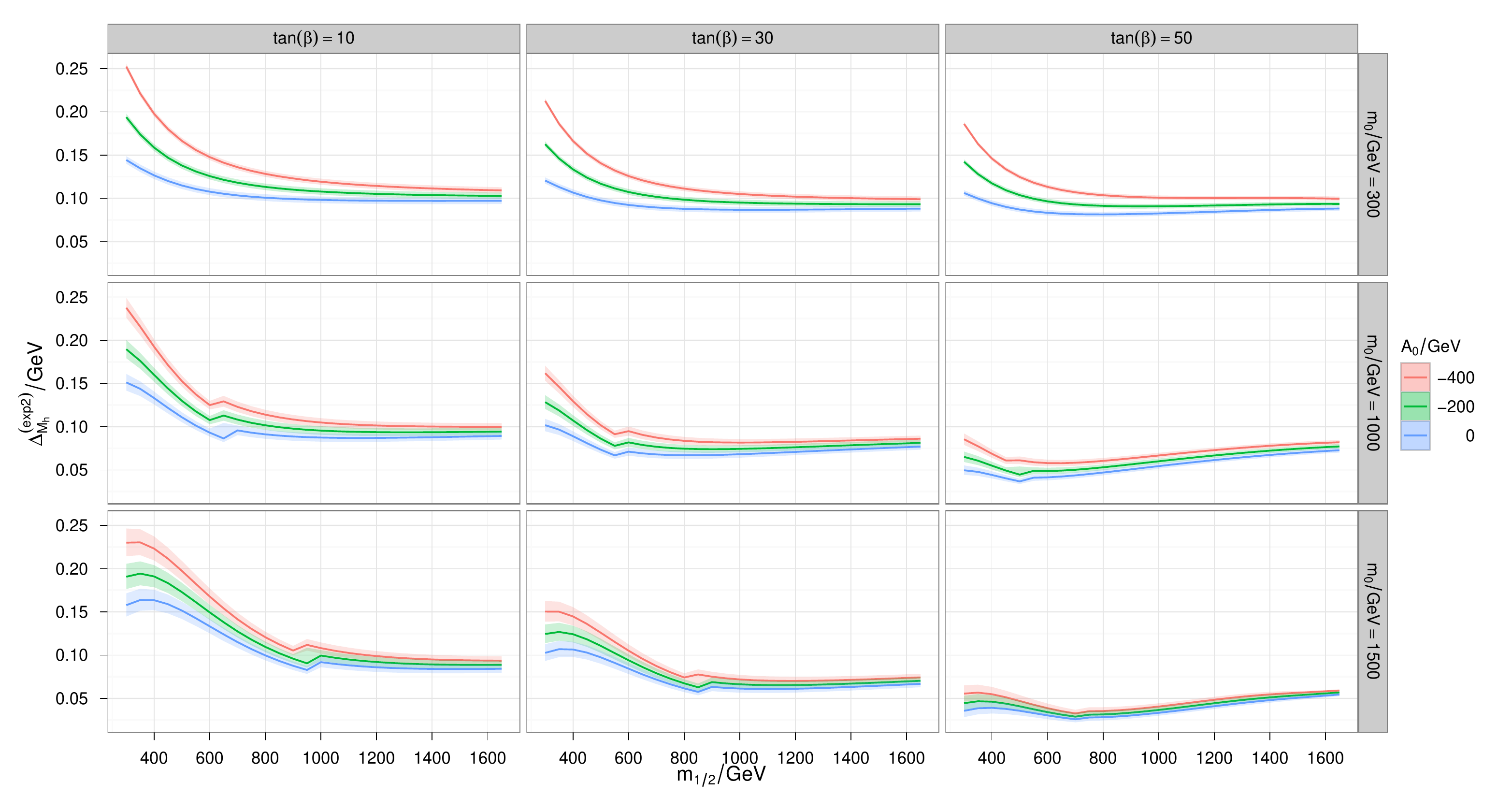}
  \caption{Deviation of the expansion in masses from the exact result
    at two loops.  The conventions are the same as in
    Fig.~\ref{fig:mhmsugra}.  The error introduced by the asymptotic
    expansions is well under control.  The little bumps in the curves
    are due to a change from one hierarchy to another.}
  \label{fig:experror}
\end{figure}

By performing asymptotic expansions, we have introduced an additional
source of uncertainty to the prediction of \mh.  With
Fig.~\ref{fig:experror}, we can analyse how large this uncertainty is.
It shows the error that we would have introduced had we made the same
approximation at the two-loop level.  We see that this error is
typically at or below $100\,$MeV, or $200\,$MeV for low values of
$m_{1/2}$ and $\tan\beta$.  Since the three-loop corrections are
smaller than the two-loop corrections, we expect the actual error due
to asymptotic expansions to be below $100\,$MeV.

\section{Conclusions}
We have presented a calculation of the \orderatasas{} corrections to
the mass \mh{} of the light Higgs boson in the \mssm.  These
contributions shift the value of \mh{} by $1-3\,$GeV, depending on the
mass spectrum of the superpartners.

The results have been implemented in the program \hthreem, which is
freely available~\cite{Kant:2010tf}.

Using our results, we improve the theoretical error significantly.
The theoretical error is now on the order of about $100\,$MeV for
light and $1\,$GeV for heavy superpartner masses.  This is comparable
to the parametric uncertainty with the top mass and \alphas.

\section{Acknowledgements}
This work was supported by the DFG through SFB/TR-9 and by the
Helmholtz Alliance ``Physics at the Terascale''.  We thank Robert
Harlander, Luminita Mihaila and Matthias Steinhauser, with whom this
work was done, Pietro Slavich and Steve Martin for providing us with
private code that implements results from~\cite{Degrassi:2001yf}
and~\cite{Martin:2005ch}, respectively, and Thomas Hahn for useful
discussions about some technical details of \FeynHiggs.

\section*{References}

\end{document}